\documentclass[fleqn,12pt,twoside]{article}
\usepackage{espcrc1}

\usepackage{graphicx}
\usepackage[figuresright]{rotating}

\def\PRL{Phys. Rev. Lett.}
\def\PRC{Phys. Rev. C}
\def\PRD{Phys. Rev. D}
\def\PLB{Phys. Lett. B}
\def\NPA{Nucl. Phys. A}
\def\NPB{Nucl. Phys. B}
\def\JPG{J. Phys. G: Nucl. Part. Phys.}
\def\IJMPE{Int. J. Mod. Phys. E}
\def\EPJA{Eur. Phys. J. A}
\def\AP{Ann. Phys. (N.Y.)}
\def\AR{Annu. Rev. Nucl. Part. Sci.}
\def\RMF{Rev. Mex. F{\'{\i}}s. S}
\def\Journal#1#2#3#4{#1 {\bf #2} (#4) #3}
\newcommand{\ba}{\begin{eqnarray}}
\newcommand{\ea}{\end{eqnarray}}

\newcommand{\AmS}{{\protect\the\textfont2
  A\kern-.1667em\lower.5ex\hbox{M}\kern-.125emS}}

\title{Flavor content of nucleon form factors in the space- and time-like region}

\author{Roelof Bijker\address{ICN-UNAM, AP 70-543, 04510 M\'exico DF, M\'exico}       
       \thanks{Supported in part by a grant from CONACyT, Mexico}}

\begin{document}

\maketitle

\begin{abstract}
I discuss a two-component model of nucleon form factors in which the external 
photon couples both to an intrinsic three-quark structure and to a meson cloud 
via vector-meson dominance, and present a simultaneous analysis of the  
electromagnetic form factors of the nucleon in the space- and time-like 
regions as well as their strangeness content.   
\end{abstract}

\section{INTRODUCTION}

The structure of the nucleon is of fundamental importance in nuclear and 
particle physics \cite{Thomas}. It has been investigated for many decades, 
from the measurement of the anomalous magnetic moment of the nucleon 
in the 1930's \cite{stern}, the radius in the 1950's \cite{hofstadter}, 
to the discovery of point-like constituents in the 1960's \cite{dis}.  

Recent experimental data on electromagnetic form factors, such as the 
momentum dependence of the ratio of electric and magnetic form factors of 
the proton from polarization transfer experiments \cite{jones,Gayou}, 
the strange form factors of the proton from parity-violating electron 
scattering (PVES) experiments \cite{Armstrong} and the ratio of electric 
to magnetic form factors in the time-like region from electron-positron 
annihilation experiments \cite{aubert}, have provided new and unexpected 
insights into the structure of the nucleon and the underlying 
dynamics of nonperturbative QCD. 

The aim of this contribution is to present a simultaneous analysis of the 
electromagnetic form factors of the nucleon in the space- and time-like 
regions as well as their strangeness content.   

\section{NUCLEON FORM FACTORS}

Electromagnetic form factors are key ingredients in the understanding of the 
internal structure of composite particles like the nucleon since they contain 
the information about the distribution of electric charge and magnetization.
The electric and magnetic form factors, $G_{E}$ and $G_{M}$, are 
obtained from the Dirac and Pauli form factors, $F_{1}$ and $F_{2}$, 
by the relations $G_E=F_1-\tau F_2$ and $G_M=F_1 + F_2$ with 
$\tau=Q^2/4 M_N^2$. 

The recent polarization transfer data for the proton form factor ratio are in excellent 
agreement with the predictions of a two-component model of the nucleon \cite{IJL} 
in which the external photon couples both to an intrinsic structure and to the 
intermediate vector mesons ($\rho$, $\omega$, $\phi$) via vector-meson dominance 
(VMD). In its original version, the Dirac form factor was attributed 
to both the intrinsic structure and the meson cloud, and the Pauli form factor 
entirely to the meson cloud. In a modified version \cite{BI}, it was shown that 
the addition of an intrinsic part to the isovector Pauli form factor as suggested 
by studies of relativistic constituent quark models in the light-front approach 
\cite{frank}, improves the results for the elecromagnetic form factors of the 
neutron considerably. 

In order to incorporate the contribution of the isocalar ($\omega$ and 
$\phi$) and isovector ($\rho$) vector mesons, it is convenient to 
introduce isoscalar and isovector Dirac and Pauli form factors. 
The isoscalar form factors contain the couplings to the $\omega$ and $\phi$ 
mesons \cite{IJL}
\ba
F_{1}^{I=0}(Q^{2}) &=& \frac{1}{2} g(Q^{2}) \left[ 
1-\beta_{\omega}-\beta_{\phi} 
+\beta_{\omega} \frac{m_{\omega }^{2}}{m_{\omega }^{2}+Q^{2}} 
+\beta_{\phi} \frac{m_{\phi}^{2}}{m_{\phi }^{2}+Q^{2}}\right] , 
\nonumber\\
F_{2}^{I=0}(Q^{2}) &=& \frac{1}{2}g(Q^{2})\left[ 
\alpha_{\omega} \frac{m_{\omega }^{2}}{m_{\omega }^{2}+Q^{2}} 
+ \alpha_{\phi} \frac{m_{\phi}^{2}}{m_{\phi}^{2}+Q^{2}}\right] ,
\label{ff}
\ea
and the isovector ones the couplings to the $\rho$ meson \cite{BI}
\ba 
F_{1}^{I=1}(Q^{2}) &=& \frac{1}{2}g(Q^{2})\left[ 1-\beta_{\rho} 
+\beta_{\rho} \frac{m_{\rho}^{2}}{m_{\rho}^{2}+Q^{2}} \right] , 
\nonumber\\ 
F_{2}^{I=1}(Q^{2}) &=& \frac{1}{2}g(Q^{2})\left[ 
\frac{\mu_{p}-\mu_{n}-1-\alpha_{\rho}}{1+\gamma Q^{2}} 
+\alpha_{\rho} \frac{m_{\rho }^{2}}{m_{\rho}^{2}+Q^{2}} \right] .
\ea
This parametrization ensures that the three-quark contribution to the 
anomalous magnetic moment is purely isovector, as given by $SU(6)$. 
The intrinsic form factor is a dipole $g(Q^{2})=(1+\gamma Q^{2})^{-2}$ 
which coincides with the form used in an algebraic treatment of the 
intrinsic three-quark structure \cite{bijker}. 
The large width of the $\rho$ meson which is crucial for the small $Q^{2}$ 
behavior of the form factors, is taken into account in the same way as in 
\cite{IJL,BI}. For small values of $Q^2$ the form factors are dominated by the 
meson dynamics, whereas for large values they satisfy the asymptotic behavior 
of p-QCD, $F_1 \sim 1/Q^4$ and $F_2 \sim 1/Q^6$ \cite{pQCD}.

\section{FLAVOR CONTENT}

The flavor content of the electromagnetic form factors of the nucleon 
can be studied by combining the nucleon's response to the electromagnetic 
and weak neutral vector currents \cite{Manohar}. 
The strange quark content is of special interest because it provides 
a direct probe of the quark-antiquark sea. It can be determined 
assuming charge symmetry and combining parity-violating asymmetries 
with measurements of the electric and magnetic form factors of the 
proton and neutron \cite{beckexp,beckth}. 

In VMD models, the strangeness content of the nucleon form factors arises through 
the coupling of the strange current to the isocalar vector mesons $\omega$ and $\phi$ 
\cite{Jaffe}. Under the assumption that the momentum dependence of the strange form 
factors is the same as that of the isoscalar ones, the strange Dirac and Pauli form 
factors are expressed as \index{}\cite{jpg}
\ba
F_{1}^{s}(Q^{2}) &=& \frac{1}{2}g(Q^{2})\left[ 
\beta_{\omega}^s \frac{m_{\omega}^{2}}{m_{\omega }^{2}+Q^{2}} 
+\beta_{\phi}^s \frac{m_{\phi}^{2}}{m_{\phi }^{2}+Q^{2}}\right] , 
\nonumber\\
F_{2}^{s}(Q^{2}) &=& \frac{1}{2}g(Q^{2})\left[ 
\alpha_{\omega}^s \frac{m_{\omega}^{2}}{m_{\omega }^{2}+Q^{2}}
+\alpha_{\phi}^s \frac{m_{\phi}^{2}}{m_{\phi }^{2}+Q^{2}}\right] .
\label{sff}
\ea
The isocalar and strange couplings appearing in Eqs.~(\ref{ff}) and (\ref{sff}) 
depend on the same nucleon-meson and current-meson couplings \cite{Jaffe} and 
are constrained by the electric charges and magnetic moments of the nucleon 
\ba
\beta_{\omega}^s/\beta_{\omega} = \alpha_{\omega}^s/\alpha_{\omega} = 
-\sqrt{6} \, \sin \epsilon/\sin(\theta_0+\epsilon) ~, 
&\hspace{1cm}& \alpha_{\omega} = \mu_p + \mu_n - 1 - \alpha_{\phi} ~, 
\nonumber\\
\beta_{\phi}^s/\beta_{\phi} = \alpha_{\phi}^s/\alpha_{\phi} =  
-\sqrt{6} \, \cos \epsilon/\cos(\theta_0+\epsilon) ~,
&\hspace{1cm}& \beta_{\phi} = -\beta_{\omega} \tan \epsilon/\tan(\theta_0+\epsilon) ~,
\label{coef2}
\ea
with $\tan \theta_0 = 1/\sqrt{2}$. The angle $\epsilon$ represents the 
deviation from the ideally mixed states  
$\left| \omega_0 \right>=(u \bar{u} + d \bar{d})/\sqrt{2}$ 
and $\left| \phi_0 \right> = s \bar{s}$. Here we use the value 
$\epsilon=0.053$ rad (or 3.0$^{\circ}$) which was determined from the 
radiative decays of the $\omega$ and $\phi$ mesons \cite{Jain}. 

\section{RESULTS}

\subsection{Space-like form factors}

In order to calculate the nucleon form factors in the two-component model, the 
five remaining coefficients, $\gamma$ from the intrinsic form factor, the isoscalar 
couplings $\beta_{\omega}$ and $\alpha_{\phi}$, and the isovector couplings 
$\beta_{\rho}$ and $\alpha_{\rho}$, are determined in a least-square fit to the 
electric and magnetic form factors of the nucleon \cite{BI,jpg}. 

\begin{figure}[htb]
\begin{minipage}[t]{80mm}
\resizebox{\textwidth}{!}{%
\includegraphics{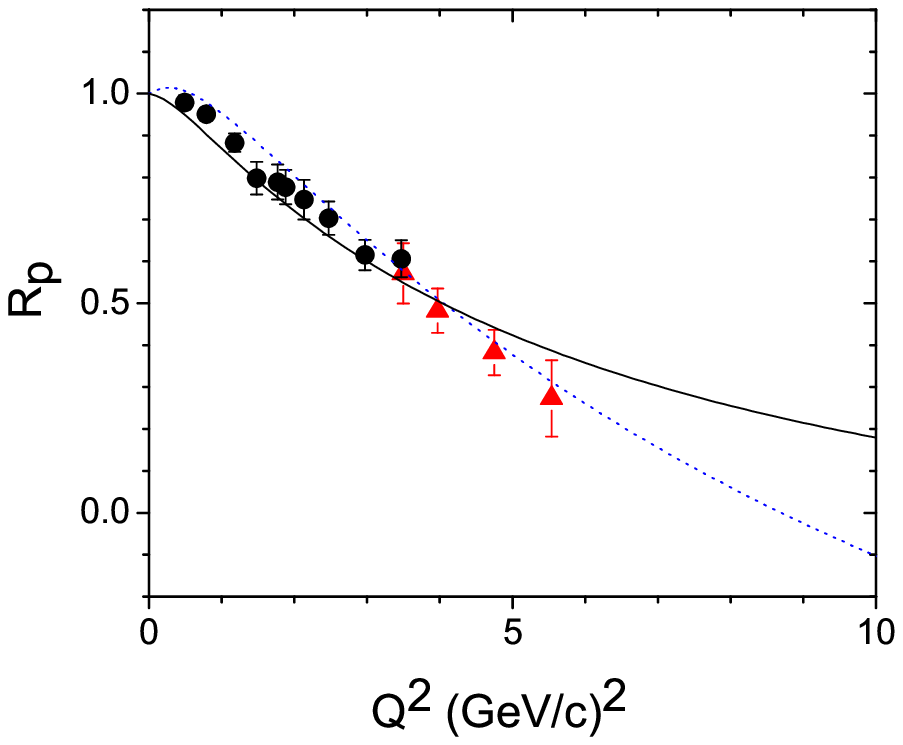}}
\end{minipage}
\hspace{\fill}
\begin{minipage}[t]{80mm}
\resizebox{\textwidth}{!}{%
\includegraphics{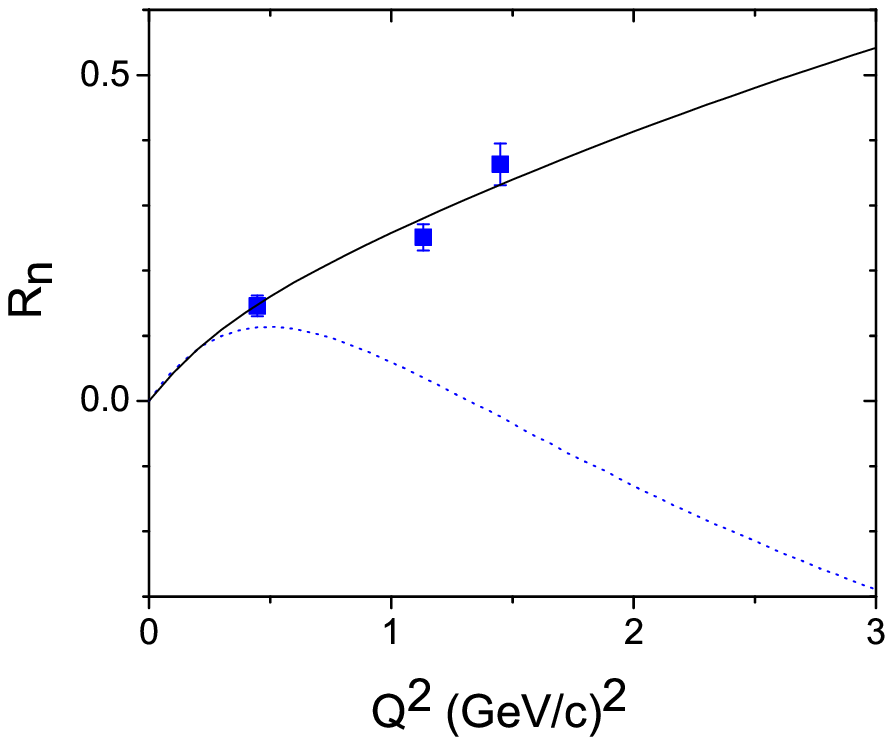}}
\end{minipage}
\caption{\small Comparison between the experimental and theoretical 
form factor ratios $R_p=\mu _{p}G_{E_{p}}/G_{M_{p}}$ (left) 
and $R_n=\mu _{n}G_{E_{n}}/G_{M_{n}}$ (right). 
The experimental data are taken from \cite{jones} (circles), 
\cite{Gayou} (triangles) and \cite{Madey} (squares).  
The solid line represents the present calculation and the 
dotted line \protect\cite{IJL}.}
\label{ffratio}
\end{figure}

Fig.~\ref{ffratio} shows the form factors ratios for the proton 
and neutron. The linear drop in 
the proton form factor ratio was predicted as early as 1973 in a VMD 
model \cite{IJL} (dotted line) and later also in a chiral soliton model 
\cite{soliton}. The experimental data for the neutron form factor ratio 
\cite{Madey} are in agreement with \cite{IJL} for small values 
of $Q^{2}$, but not so for higher values of $Q^2$. The present calculation 
(solid line) is in good agreement with the data, especially for the neutron. 

\subsection{Strange form factors}

\begin{figure}[htb]
\begin{minipage}[t]{80mm}
\resizebox{\textwidth}{!}{%
\includegraphics{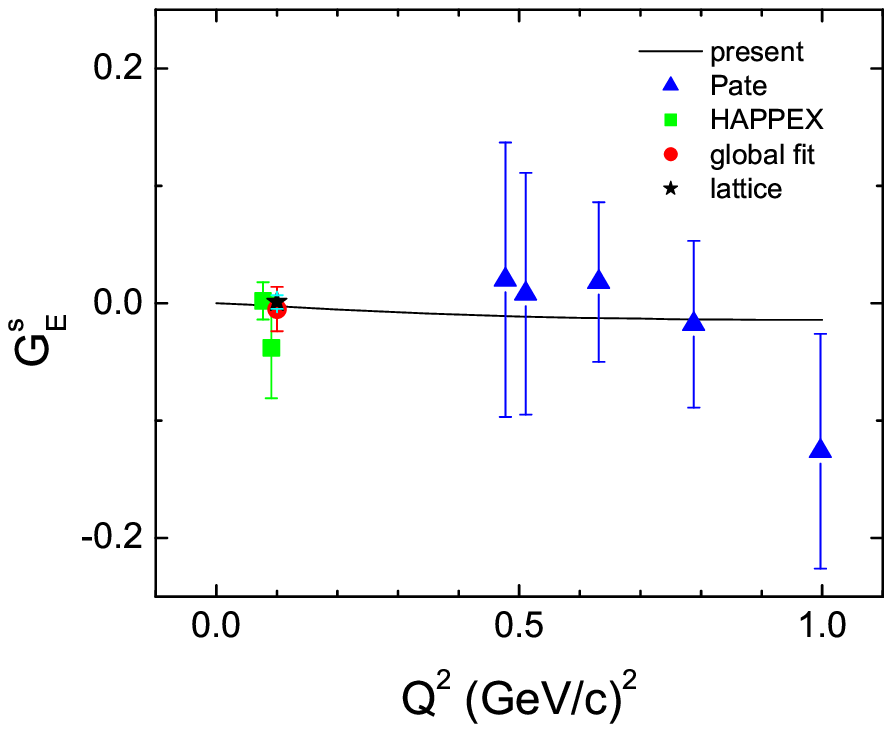}} 
\end{minipage}
\hspace{\fill}
\begin{minipage}[t]{80mm}
\resizebox{\textwidth}{!}{%
\includegraphics{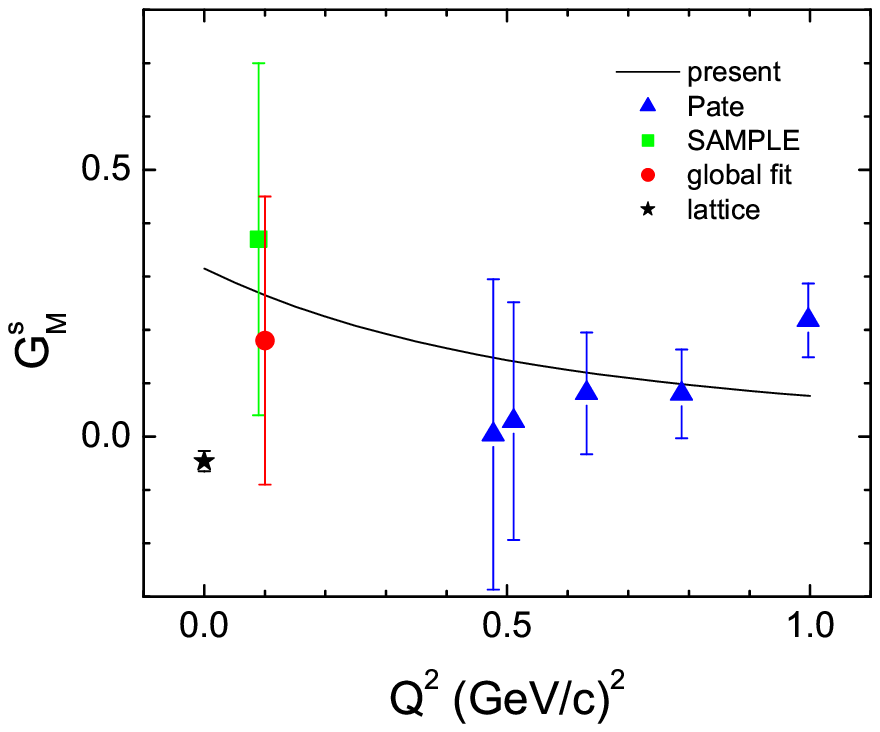}}
\end{minipage}
\caption{\small Comparison between theoretical and experimental values 
of the strange electric (left) and magnetic (right) form factor. 
The experimental values are taken from \cite{AniolPRL,Acha,Spayde} (squares),  
\cite{Acha} (circles) and \cite{Pate} (triangles). 
The lattice results are taken from \cite{lattice} (stars).}
\label{GEMs}
\end{figure}

The strange form factors are obtained by combining Eqs.~(\ref{sff}) 
and (\ref{coef2}) with the fitted values of $\beta_{\omega}$ and 
$\alpha_{\phi}$ \cite{jpg,milos}. Figure~\ref{GEMs} shows the strange 
electric and magnetic form factors as a function of $Q^2$. 
The qualitative features of these form factors can be understood 
easily in the limit of ideally mixed mesons, {\em i.e.} zero mixing angle 
$\epsilon=0^{\circ}$ (in comparison with the value of $\epsilon=3.0^{\circ}$ 
used in the present calculations). In this limit, the Dirac form factor 
vanishes identically and the Pauli form factor only depends 
on the tensor coupling $\alpha_{\phi}^s$. 
Therefore, the strange magnetic form factor $G_M^s=F_2^s$ drops as 
$1/Q^6$ and has the same sign as $\alpha_{\phi}^s$ (positive), 
whereas for small values of $Q^2$ the strange electric form factor  
$G_E^s=-\tau F_2^s$ is suppressed by $\tau=Q^2/4M_N^2$. 

The theoretical values for $G_E^s$ are small and negative, in agreement 
with the experimental results of the HAPPEX Collaboration in which $G_E^s$ 
was determined in PVES from $^{4}$He \cite{AniolPRL,Acha}. 
The values of $G_M^s$ are positive, since they dominated by the 
contribution from the Pauli form factor. The sign and magnitude are in 
agreement with the experimental result from SAMPLE \cite{Spayde}. 
A global fit of all measurements at $Q^2=0.1$ (GeV/c)$^2$ 
gives $G_M^s=0.18 \pm 0.27$ \cite{Acha}.  
The other experimental values of $G_E^s$ and $G_M^s$ in Fig.~\ref{GEMs} 
for $0.4 < Q^2 < 1.0$ (GeV/c)$^2$ were obtained \cite{Pate} by combining 
the (anti)neutrino scattering data \cite{Ahrens} with the parity-violating 
asymmetries \cite{Armstrong,Aniol04}. The strange magnetic moment is 
calculated to be $\mu_s= 0.315$ $\mu_N$, whose sign is in 
contradiction with most theoretical calculations \cite{beckexp,beckth}, 
but in agreement with the available experimental data \cite{Acha,Spayde,AniolPLB}. 
Recent lattice calculations of the strange magnetic moment give a small 
negative value $\mu_s= -0.046 \pm 0.019$ $\mu_N$ \cite{lattice}. 
 
Most available data on strange form factors are for linear combinations 
of electric and magnetic form factors $G_E^s+\eta G_M^s$ 
\cite{Armstrong,Acha,Aniol04,AniolPLB,Maas}. An analysis of these data 
shows a good overall agreement with the theoretical values of the 
two-component model \cite{jpg,milos}. 

\subsection{Time-like form factors}

For a global understanding of the structure of the nucleon it is important 
to study the nucleon form factors in the time-like region as well 
\cite{hammer,dub,TG}. Whereas in the space-like (SL) region 
($Q^2 > 0$) the electromagnetic form factors can be studied through electron 
scattering, in the time-like (TL) region ($q^2=-Q^2 > 0$) they can be measured 
through the creation or annihilation of a nucleon-antinucleon pair. 
SL nucleon form factors are real because of the hermiticity of the electromagnetic 
interaction, while TL form factors are complex. Theoretically, they are related 
by analytic continuation $Q^2=-q^2 \rightarrow q^2\exp{(-i\pi)}$:  
$F^{(SL)}(Q^2) \rightarrow F^{(TL)}(q^2)$. 
The analytic structure of the form factors leads to a rigorous 
prescription of their asymptotic behavior via the Phragm\`en-Lindel\"of theorem 
\cite{TGakh}  
\ba
\lim_{Q^2 \rightarrow \infty} F^{(SL)}(Q^2) = 
\lim_{q^2 \rightarrow \infty} F^{(TL)}(q^2) ,
\ea
implying that in the asymptotic limit the imaginary part of the TL form 
factors vanishes, whereas the real parts of the TL and SL form factors coincide. 
In previous studies of TL form factors in the 
two-component model \cite{BI,wan} a phase was added to the intrinsic part in 
order to move the singularity at $q^2=1/\gamma$ from the real axis. This extra 
phase has however the disadvantage that the Phragm\`en-Lindel\"of theorem 
is no longer satisfied \cite{TGakh}. 
 
\begin{figure}[htb]
\begin{minipage}[t]{80mm}
\resizebox{\textwidth}{!}{%
\includegraphics{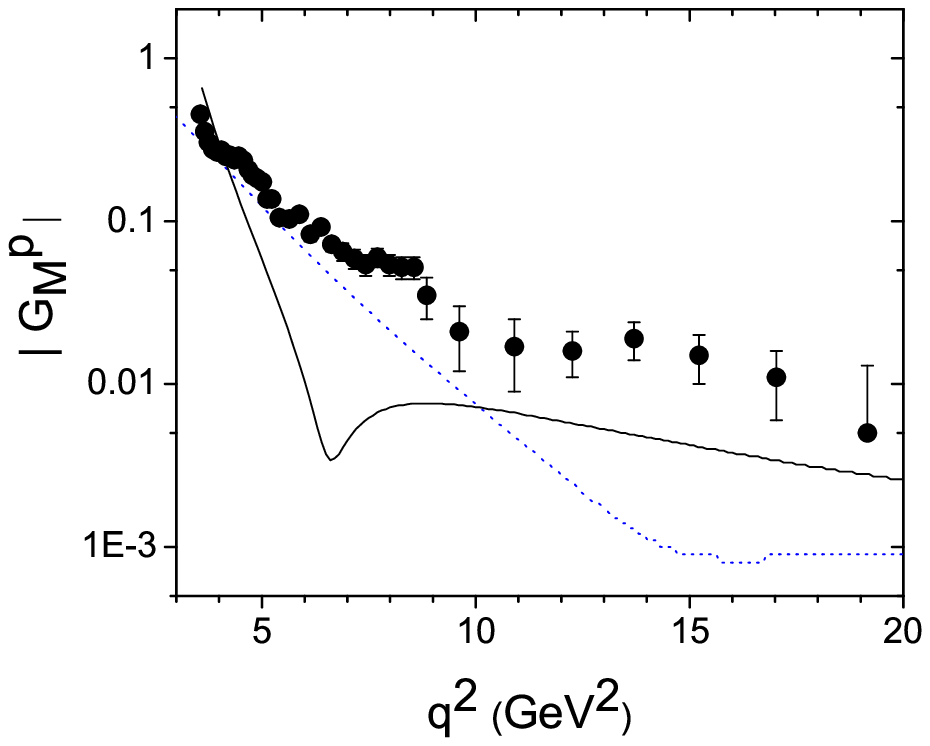}} 
\end{minipage}
\hspace{\fill}
\begin{minipage}[t]{80mm}
\resizebox{\textwidth}{!}{%
\includegraphics{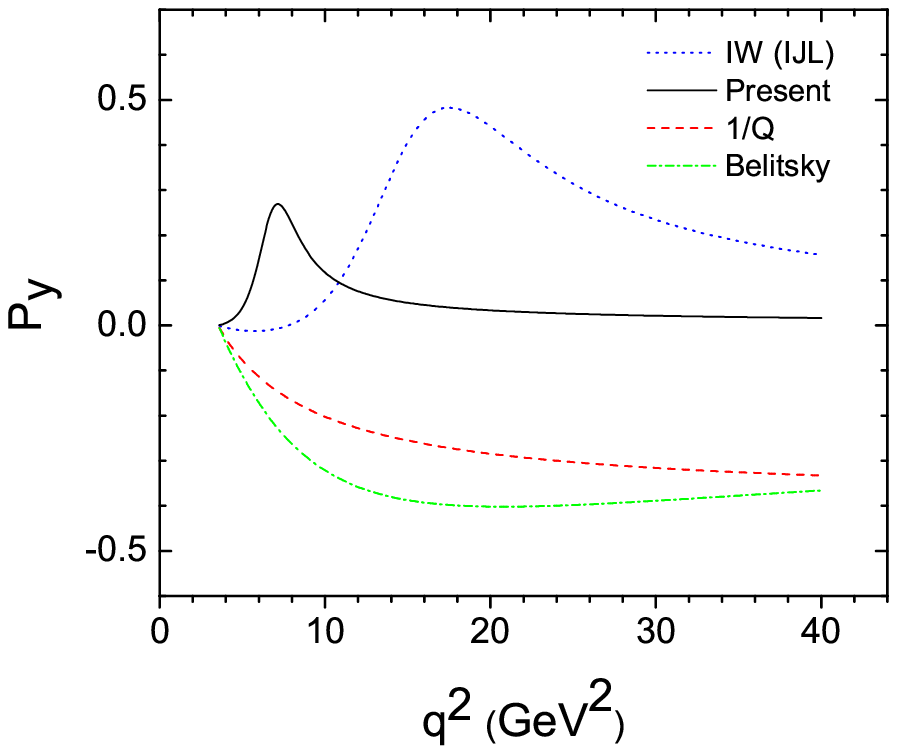}}
\end{minipage}
\caption{\small (left) Comparison between theoretical and experimental values 
of the time-like proton magnetic form factor $|G_{M_p}|$. 
The experimental values are taken from \cite{aubert} under the assumption  
$|G_{E_p}|=|G_{M_p}|$. The solid line represents the present calculation 
and the dotted line \protect\cite{wan}. (right) Comparison of different 
predictions for $P_y$. The solid line represents the present calculation, 
and the others are taken from \cite{BCHH}.}
\label{ptime}
\end{figure}

Fig.~\ref{ptime} shows that the present calculation agrees with the recent 
BABAR data \cite{aubert} close to threshold ($q^2 = 4M_N^2$) and for large 
values of $q^2$, but shows serious discrepancies for intermediate values. 
At the moment, the available experimental information does not allow to 
separate the contributions from the electric and magnetic form factors, 
nor to measure their relative phase. In the extraction of $|G_{M_p}|$ from 
the data it was assumed that $|G_{E_p}|=|G_{M_p}|$ 
which is true at threshold, but not in general. 

Polarization observables are sensitive probes of different models of the nucleon,  
as can be seen in the second panel of Fig.~\ref{ptime} which shows a comparison 
of different theoretical predictions for $P_y \propto \mbox{Im} (G_E^* G_M)$. 
The present calculation is the only one for which $P_y$ vanishes in the asymptotic 
region, as required by the Phragm\`en-Lindel\"of theorem. 

\section{SUMMARY AND CONCLUSIONS}

In this contribution, I presented the results of a simultaneous study 
of the nucleon form factors in the space- and time-like regions as well 
as their strangeness content. The analysis was carried out in a  
VMD approach in which the two-component model of Bijker 
and Iachello for the electromagnetic nucleon form factors \cite{BI} 
is combined with the method proposed by Jaffe to determine the strangeness 
content via the coupling of the strange current to the $\phi$ and $\omega$ 
mesons \cite{Jaffe}. The strange couplings are completely fixed by 
the electromagnetic form factors of the proton and neutron. 

A comparison with the available experimental data on the space-like form 
factors and their strange quark content shows that the present approach 
provides a simultaneous and consistent description of the electromagnetic 
and weak vector form factors of the nucleon. The strangeness 
contribution to the charge and magnetization distributions is of the 
order of a few percent \cite{milos}. Future experiments on PVES to backward 
angles and neutrino scattering will make it possible to determine the 
contributions of the different quark flavors to the electric, magnetic 
and axial form factors, and thus to gain new insights into the complex 
structure of the nucleon. 

However, in the time-like region there are some serious discrepancies. 
The present results point once again to the inconsistency between space-like 
and time-like data already noted in \cite{BI,hammer,wan}. 
In the extraction of the experimental values of the proton magnetic form 
factor it is assumed that $|G_{E_{p}}|=|G_{M_{p}}|$ which is true 
at threshold, but not for $q^2 > 4M_N^2$. For the neutron, the electric 
form factor is assumed to be zero $|G_{E_{n}}|=0$. A remeasurement of 
the time-like data in which the contributions of the electric and magnetic 
form factors are separated, as well as a measurement of the polarization 
observables and a determination of the relative phase of electric and 
magnetic form factors may help to resolve this inconsistency.


\begin{thebibliography}{99}

\bibitem{Thomas}
A.W. Thomas and W. Weise, 
{\em The Structure of the Nucleon}, 
(Wiley-VCH, Berlin, 2001). 

\bibitem{stern}
I. Estermann, R. Frisch and O. Stern, 
Nature {\bf 132} (1933) 169.

\bibitem{hofstadter}
R. Hofstadter,  
Annu. Rev. Nucl. Sci. {\bf 7} (1957) 231.

\bibitem{dis}
J.I. Friedman and H.W. Kendall, 
Annu. Rev. Nucl. Sci. {\bf 22} (1972) 203.

\bibitem{jones}  
M.K. Jones \textit{et al.}, 
\Journal{\PRL}{84}{1398}{2000}; \\ 
V. Punjabi {\em et al.}, 
\Journal{\PRC}{71}{055202}{2005}. 

\bibitem{Gayou} 
O. Gayou {\em et al.}, 
\Journal{\PRL}{88}{092391}{2002}.

\bibitem{Armstrong}
D.S. Armstrong {\em et al.},  
\Journal{\PRL}{95}{092001}{2005}.   

\bibitem{aubert}
B. Aubert {\em et al.}, 
\Journal{\PRD}{73}{012005}{2006}. 

\bibitem{IJL}  
F. Iachello, A.D. Jackson and A. Lande, 
\Journal{\PLB}{43}{191}{1973}.

\bibitem{BI}
R. Bijker and F. Iachello, 
\Journal{\PRC}{69}{068201}{2004}; \\
R. Bijker, \Journal{\RMF}{52}{17}{2006} 
[arXiv:nucl-th/0502050]. 

\bibitem{frank}  
M.R. Frank, B.K. Jennings and G.A. Miller, 
\Journal{\PRC}{54}{920}{1996}; \\
E. Pace, G. Salm\`e, F. Cardarelli and S. Simula,  
\Journal{\NPA}{666}{33c}{2000}.

\bibitem{bijker}  
R. Bijker, F. Iachello and A. Leviatan, 
\Journal{\AP}{236}{69}{1994}. 

\bibitem{pQCD}
G.P. Lepage and S.J. Brodsky, 
\Journal{\PRL}{43}{545}{1979}; \\
G.P. Lepage and S.J. Brodsky, 
\Journal{\PRD}{22}{2157}{1980}. 

\bibitem{Manohar}
D.B. Kaplan and A. Manohar, \Journal{\NPB}{310}{527}{1988}; \\
R.D. McKeown, \Journal{\PLB}{219}{140}{1989}; \\
D.H. Beck, \Journal{\PRD}{39}{3248}{1989}. 

\bibitem{beckexp}
D.H. Beck and R.D. McKeown, 
\Journal{\AR}{51}{189}{2001}. 

\bibitem{beckth}
D.H. Beck and B.R. Holstein, 
\Journal{\IJMPE}{10}{1}{2001}. 

\bibitem{Jaffe}
R.L. Jaffe, 
\Journal{\PLB}{229}{275}{1989}. 

\bibitem{jpg}
R. Bijker, 
\Journal{\JPG}{32}{L49}{2006} [arXiv:nucl-th/0511060].

\bibitem{Jain}
P. Jain {\em et al.},     
\Journal{\PRD}{37}{3252}{1988}.  

\bibitem{Madey} 
R. Madey {\em et al.},
\Journal{\PRL}{91}{122002}{2003}.

\bibitem{soliton} 
G. Holzwarth, 
Z. Phys. A \textbf{356} (1996) 339.

\bibitem{milos}
R. Bijker, 
arXiv:nucl-th/0607058. 

\bibitem{AniolPRL}
K.A. Aniol {\em et al.},  
\Journal{\PRL}{96}{022003}{2006}.  

\bibitem{Acha}
A. Acha {\em et al.}, arXiv:nucl-ex/0609002.

\bibitem{Spayde}
D.T. Spayde {\em et al.},  
\Journal{\PLB}{583}{79}{2004}; \\ 
E.J. Beise, M.L. Pitt and D.T. Spayde, 
Prog. Part. Nucl. Phys. {\bf 54} (2005) 289. 

\bibitem{Pate}
S.F. Pate, \Journal{\PRL}{92}{082002}{2004}; \\  
S.F. Pate {\em et al.}, arXiv:hep-ex/0512032.

\bibitem{lattice}
D.B. Leinweber {\em et al.},    
\Journal{\PRL}{94}{212001}{2005}; \\ 
D.B. Leinweber {\em et al.},    
\Journal{\PRL}{97}{022001}{2006}.

\bibitem{Ahrens}
L.A. Ahrens {\em et al.},  
\Journal{\PRD}{35}{785}{1987}.  
 
\bibitem{Aniol04}
K.A. Aniol {\em et al.},  
\Journal{\PRC}{69}{065501}{2004}. 

\bibitem{AniolPLB}
K.A. Aniol {\em et al.}, 
\Journal{\PLB}{635}{275}{2006}.

\bibitem{Maas}
F.E. Maas {\em et al.}, \Journal{\PRL}{93}{022002}{2004}; \\ 
F.E. Maas {\em et al.}, \Journal{\PRL}{94}{152001}{2005}. 

\bibitem{hammer}  
H.-W. Hammer, U.-G. Meissner and D. Drechsel, 
\Journal{\PLB}{385}{343}{1996}.

\bibitem{dub}
S. Dubni\v{c}ka {\em et al.}, 
\Journal{\JPG}{29}{405}{2003}.

\bibitem{TG}
E. Tomasi-Gustafsson {\em et al.}, 
\Journal{\EPJA}{24}{419}{2005}.

\bibitem{TGakh}
E. Tomasi-Gustafsson and G.I. Gakh, 
\Journal{\EPJA}{26}{285}{2005}.

\bibitem{wan} 
F. Iachello and Q. Wan, 
\Journal{\PRC}{69}{055204}{2004}.

\bibitem{BCHH}
S.J. Brodsky {\em et al.},  
\Journal{\PRD}{69}{054022}{2004}.

\end{thebibliography}
\end{document}